\documentclass[twocolumn]{./jpsj2}

\usepackage{amsmath}
\usepackage{txfonts}
\usepackage{bm}

\newcommand{\sub}[1]{$_{\mathrm {#1}}$}
\newcommand{\subm}[1]{_{\mathrm {#1}}}
\newcommand{\sps}[1]{$^{\mathrm {#1}}$}

\newcommand{\etal}{\textit{et~al.}}
\newcommand{\degc}{$^{\circ}$C}

\newcommand{\cro}[1]{{#1}Cu\sub{3}Ru\sub{4}O\sub{12}}
\newcommand{\acro}{\cro{\textit{A}}}
\newcommand{\ccro}{\cro{Ca}}
\newcommand{\ncro}{\cro{Na}}
\newcommand{\lcro}{\cro{La}}
\newcommand{\nc}{Na\sub{0.5}Ca\sub{0.5}}
\newcommand{\cl}{Ca\sub{0.5}La\sub{0.5}}
\newcommand{\nccro}{\cro{\nc}}
\newcommand{\clcro}{\cro{\cl}}

\newcommand{\AcBo}[2]{{#1}Cu\sub{3}{#2}\sub{4}O\sub{12}}
\newcommand{\acbo}{\AcBo{\textit{A}}{\textit{B}}}
\newcommand{\ccbo}{\AcBo{Ca}{\textit{B}}}
\newcommand{\actro}{\textit{A}Cu\sub{3}Ti\sub{4-\textit{x}}Ru\sub{\textit{x}}O\sub{12}}

\newcommand{\Akakko}[1]{(\textit{A} = {}{#1})}
\newcommand{\Aall}{\Akakko{Na, \nc{}, Ca, \cl{}, La}}

\newcommand{\CuKa}{Cu$K_{\alpha1}$}

\title{%
Heavy-Mass Behavior of Ordered Perovskites \textit{A}Cu\sub{3}Ru\sub{4}O\sub{12} (\textit{A} = Na, Ca, La)
}

\author{%
Soutarou~Tanaka\thanks{E-mail address: stanaka@scphys.kyoto-u.ac.jp},
Nobuhiro~Shimazui, Hiroshi~Takatsu, Shingo~Yonezawa, and~Yoshiteru Maeno
}

\inst{%
Department of Physics, Graduate School of Science, Kyoto University, Kyoto 606-8502, Japan\\
}

\abst{%
We synthesized \acro{} \Aall{} and measured their DC magnetization, 
AC susceptibility, specific heat, and resistivity, 
in order to investigate the effects of the hetero-valent substitution.
A broad peak in the DC magnetization around 200~K was observed \textit{only} in \ccro{},
suggesting the Kondo effect due to localized Cu\sps{2+} ions.
However, the electronic specific heat coefficients $\gamma$ exhibit large values 
not only for \ccro{} but also for all the other samples.
Moreover, the Wilson ratio and the Kadowaki-Woods ratio of our samples 
are all similar to the values of other heavy-fermion compounds.
These results question the Kondo effect as the dominant origin of the mass enhancement, 
and rather indicate the importance of correlations among itinerant Ru electrons.
}

\kword{%
transition-metal oxide, ordered perovskite, ruthenium oxide, heavy fermion, electron correlation, Kondo effect
}

\recdate{\today}

\begin{document}
\maketitle

\section{Introduction}

\begin{figure}[tb]
\begin{center}
\includegraphics[width=9.0cm]{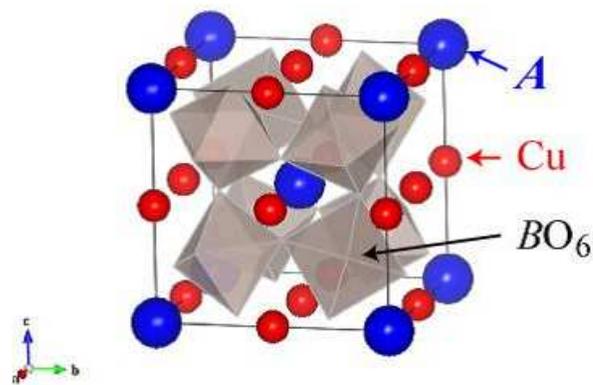}
\caption{(Color online)
Crystal structure of \acbo{}. 
Each of the octahedra is \textit{B}O\sub{6}, 
in which the \textit{B} ion is located at the center and the O ions occupy the corners.
Each O ion is shared by two octahedra.
The small red spheres represent Cu atoms and the large blue spheres represent \textit{A} atoms.
The figure is generated using the program "VESTA"\cite{VESTA200806}.
\label{fig:f1}}
\end{center}
\end{figure}

\textit{A}-site ordered perovskite oxides \acbo{} 
(where \textit{A} is alkaline-earth metals, rare-earth metals or some other elements, 
and \textit{B} is transition metals) have recently been extensively studied,
because rich varieties of electronic and magnetic properties are realized 
by various combinations of \textit{A} and \textit{B} ions\cite{Vasilev}.
For example, insulating \ccbo{}
(\textit{B} = Ti\cite{Subra2000JSolStaChem,Rami2000SolStaCommun,Homes2001Science,Subra2002SolStaSci,Kim2002SolStaCommun}, 
Ge\cite{Ozaki1977ActaCryst,Shiraki2007PRB}, 
Sn\cite{Shiraki2007PRB}) exhibit antiferromagnetism for \textit{B} = Ti\cite{Kim2002SolStaCommun} 
and ferromagnetism for \textit{B} = Ge and Sn.
Such variation is crucially decided by the \textit{d} electrons in the \textit{B} ion shells, 
rather than the ionic radii of the non-magnetic \textit{B} ions\cite{Shiraki2007PRB}.
Semiconducting \ccbo{}
(\textit{B} = Mn\cite{Bochu1979JSolStaChem,Zeng1999PRL,Liu2006JMatChem}, 
Fe\cite{Xiang2007APL,Yamada2008ACIE}) exhibit giant magnetoresistance originating from their ferrimagnetic ordered states.
Metallic \ccbo{ }(\textit{B} = V\cite{Kadyrova2003Dok,Shiraki2008JPSJ}, 
Cr\cite{Subra2003PhysChemSolids,Xiang2007InorgChem}, 
Co, 
Ru\cite{Labeau1980JSolStaChem,Ebb2002JSolStaChem,Koba2004JPSJ,Rami2004SolStaCommun,Tran2006PRB,Kato2007JMagnMagnMater,Kato2007JPhysChemSolids,Xiang2007PRB,Krimmel2008PRB}) 
exhibit Pauli paramagnetic behavior.
The crystal structure of \acbo{} is depicted in Fig.~\ref{fig:f1}.
The \textit{A} and Cu ions order in the \textit{A}-site of the perovskite \textit{AB}O\sub{3} in the ratio of one to three.
The \textit{B} and O ions form \textit{B}O\sub{6} octahedra that share the oxygen atoms at the corner 
and the octahedra are in many cases severely tilted.

Among these \textit{A}-site ordered perovskite oxides, 
\acro{}\cite{Labeau1980JSolStaChem} (\textit{A} = Na, Ca, La) has attracted much attention, 
because of its metallic conductivity with a heavy effective mass 
reported by Ramirez~\etal\cite{Rami2004SolStaCommun}{} and Kobayashi~\etal\cite{Koba2004JPSJ}
These two groups disclosed the metal-insulator transition 
in \actro{} (\textit{A} = Na, Ca, La).
Ramirez~\etal\cite{Rami2004SolStaCommun} additionally compared the resistivity and the specific heat 
of the \textit{A}-site substituted systems.
In \ccro{}, Kobayashi~\etal\cite{Koba2004JPSJ} found magnetic behavior 
ascribable to the lattice Kondo effect 
between the localized Cu\sps{2+} ions with $s=1/2$ spins 
and the itinerant electrons originating from the Ru 4\textit{d} orbitals.
They further argued that the mass enhancement is attributable to the Kondo effect.
In contrast, on the basis of the band structure calculation 
using exchange correlation functional, 
Xiang~\etal\cite{Xiang2007PRB} recently suggested that 
both the Ru 4\textit{d} band itself and a mixed band 
where the Ru 4\textit{d} band hybridizes with the nearly localized Cu 3\textit{d} band 
might contribute to the heavy-mass behavior.
Quite recently, Krimmel~\etal\cite{Krimmel2008PRB} reported 
the possibility of non-Fermi-liquid behavior below 2~K in \ccro{} 
based on the specific heat and the spin-lattice relaxation rate.


At present, it is not conclusive whether or not the Kondo effect is dominant 
or ever effective for the observed mass enhancement.
Thus, it is useful to examine compounds 
in which the formal valence of Cu is shifted from 2+ in order to weaken the Kondo effect.
This can be done by substituting Ca\sps{2+} ions 
in order not to introduce direct disorder at the Cu or Ru site.
We compare the physical properties of \ccro{} with those of \ncro{} and \lcro{}
to examine the Kondo scenario.
(Note that the formal valence of Cu would be equal to 2+ for Ca\sps{2+} 
and deviate from 2+ as the valence of \textit{A}-site varies from 2+.)
Kobayashi~\etal~\cite{Koba2004JPSJ} controlled the heavy-mass behavior 
by substituting Mn for the Cu site or Ti for the Ru site, 
but these substitutions introduce much disorder at the Cu or Ru site, 
which are considered responsible for the electric conductivity.
Because the Cu and Ru sites are directly involved in the Kondo mechanism, 
the \textit{A}-site substitution without disorder at the Cu and Ru sites may be more useful 
to examine the origin of the mass enhancement of \ccro{}.


The main discussion in this paper is on the dominant origin 
of the mass enhancement of \acro{} (\textit{A} = Na, Ca, La), 
based on the measurements of the susceptibility, specific heat, and resistivity.
The electronic specific heat coefficients $\gamma$ of all our samples 
are relatively large compared with 
those of transition-metal oxides without strong correlations 
but comparable to the ruthenate superconductor 
Sr\sub{2}RuO\sub{4}~\cite{Maeno1994Nature,Maeno1997JPSJ} with strong correlations.
Moreover the effective mass of \lcro{} is even heavier than that of \ccro{}.
This result indicates that 
the Kondo effect does not play a dominant role for the mass enhancement of \ccro{}.
Otherwise, $\gamma$ of \ccro{} would be the largest.
For all our samples, the Wilson ratio 
and the Kadowaki-Woods ratio correspond well to other heavy-fermion compounds.
These ratios provide additional evidence that the Kondo effect is not the dominant origin.
Nevertheless, the susceptibility-maximum in \ccro{} implies the presence of the Kondo effect.
We propose that 
the correlations among itinerant Ru electrons are dominant for the mass enhancement 
and the Kondo effect provides a minor contribution to the mass enhancement.

\section{Experimental}

\begin{figure}
\begin{center}
\includegraphics[width=5.0cm,bb=15 15 582 819,angle=270]{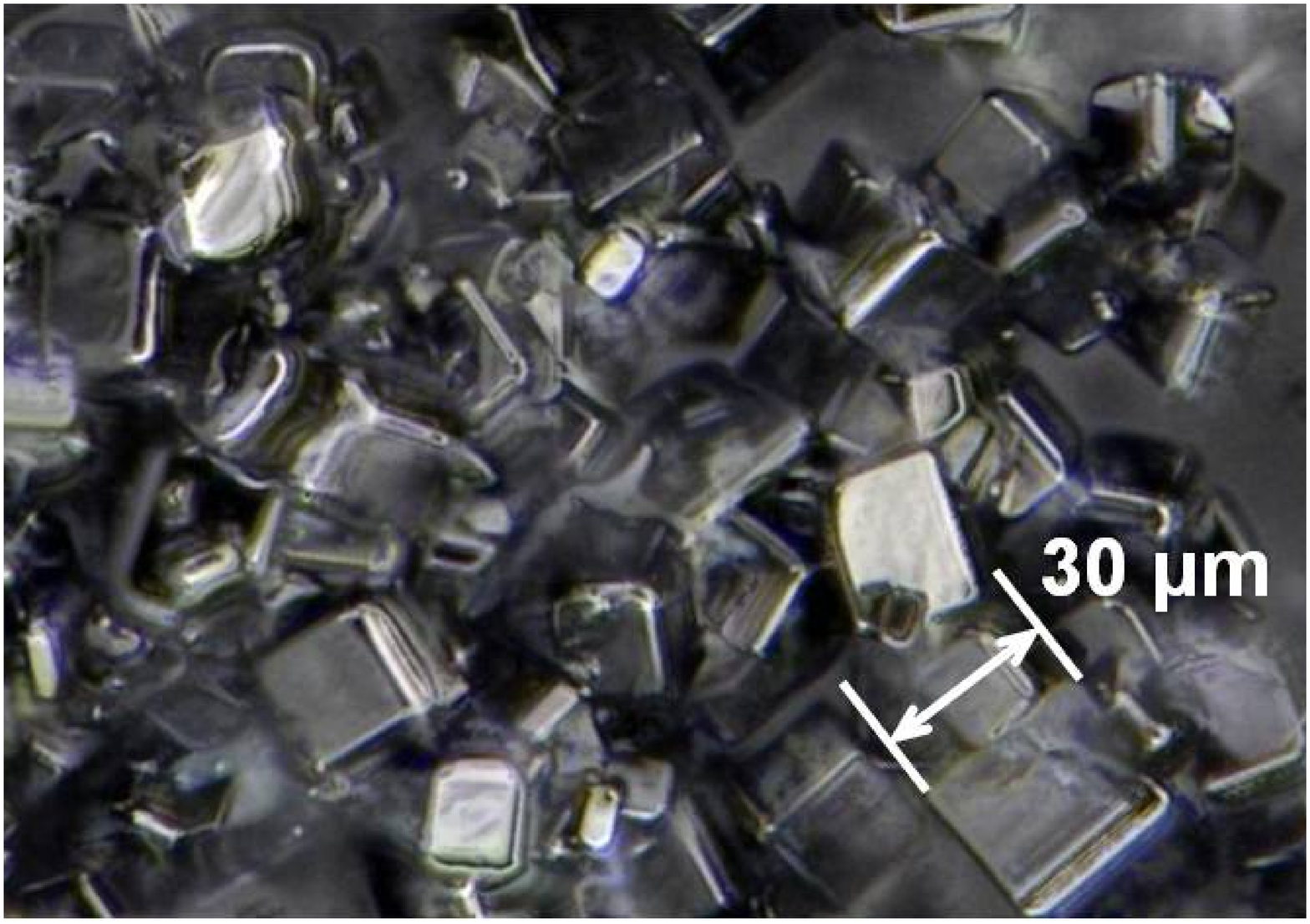}
\caption{(Color online)
Optical microscope image of single crystals of \ncro{}.
Obtained crystals are black and have a cubic shape with a side length of up to about 50~$\mu${}m.
\label{fig:f2}}
\end{center}
\end{figure}

\begin{figure}
\begin{center}
\includegraphics[width=6.0cm,angle=270]{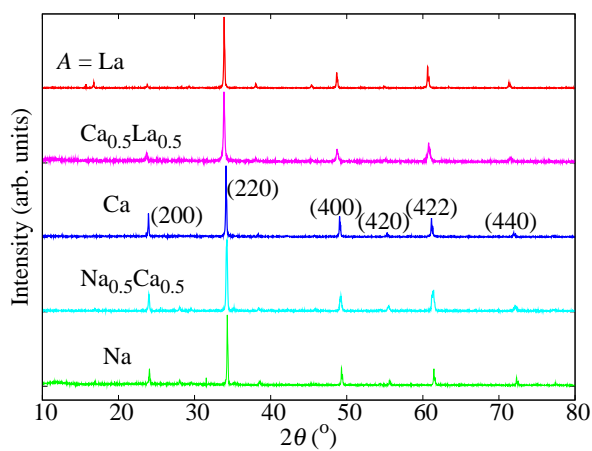}
\caption{(Color online)
Powder X-ray diffraction patterns of \acro{} at room temperature, 
demonstrating nearly single-phase samples with the structure shown in Fig.~1.
\label{fig:f3}}
\end{center}
\end{figure}

Samples of \acro{} were synthesized with a conventional solid state reaction.
Well ground stoichiometric mixtures of powders of 
Na\sub{2}CO\sub{3}, CaCO\sub{3}, La\sub{2}O\sub{3}, CuO, and RuO\sub{2} 
were pressed into pellets, placed in alumina crucibles, 
and heated and kept at 1000\degc{} for about 1 day.
Only for \ncro{}, we used 5\%{} less RuO\sub{2} 
since unreacted RuO\sub{2} was detected when we used a stoichiometric mixture.
Then they were reground, pelletized, and heated, 
and kept at 1000\degc{} for about 1 day.
By repeating this procedure for 1--3 times, 
we obtained the polycrystalline samples of \nccro{}, \ccro{}, \clcro{}, and \lcro{}.
It is difficult to synthesize \ncro{} because of its lower melting point 851\degc{}.
We first heated the ground mixture at 750\degc{}, 
and the reground mixture was pelletized and heated at 900\degc{}.
Then, the sample consisting of fine single crystals shown in Fig.~\ref{fig:f2} was obtained.
To characterize these samples, we used powder X-ray diffractometry with the \CuKa{} radiation.
As shown in Fig.~\ref{fig:f3}, we confirm that 
the impurity peaks of CuO and RuO\sub{2} are estimated less than 4\%{} from the X-ray diffraction patterns, 
indicating almost single-phase samples.
The cubic lattice parameter $a$ systematically increases 
from 0.739~nm for \textit{A} = Na to 0.748~nm for \textit{A} = La (See Table~\ref{table:t1}).

The DC susceptibility was measured from 1.8~K to 350~K 
with a commercial SQUID magnetometer (Quantum Design, model MPMS), 
and the specific heat from 2~K to 300~K 
with a commercial calorimeter (Quantum Design, model PPMS).
The resistivity was measured from 2~K to 350~K 
using a standard four-probe method with DC or AC current.
We measured the resistivity on at least three samples 
from each batch and confirmed their reproducibility.
The AC susceptibility is measured by a mutual inductance method 
down to 20~mK using a {}$^3$He-{}$^4$He dilution refrigerator 
(Cryoconcept, model DR-JT-S-100-10).

\section{Results}

\begin{figure}
\begin{center}
\includegraphics[width=9.0cm]{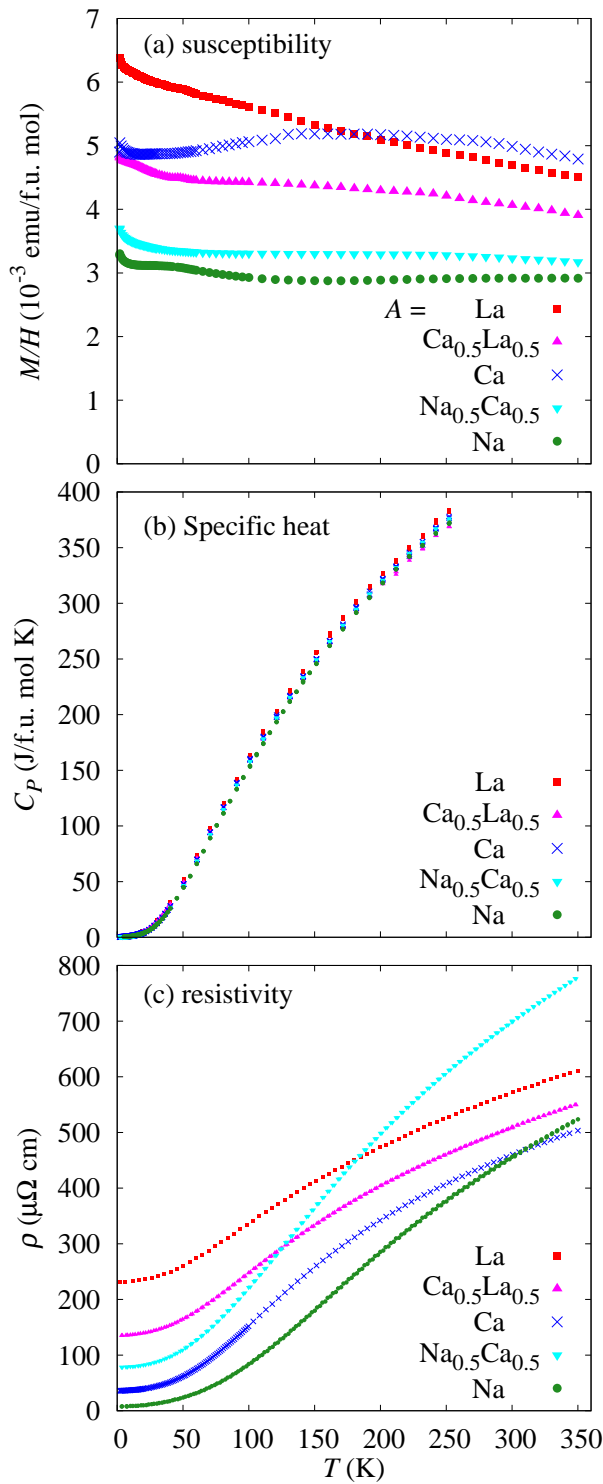}
\caption{(Color online)
Temperature dependence of %
(a) the susceptibility $\chi=M/H$ under 10~kOe, 
(b) the specific heat $C_P$, 
and (c) the resistivity $\rho$ of \acro{}.
\label{fig:f4}}
\end{center}
\end{figure}


Figure~\ref{fig:f4}(a) displays the measured DC susceptibility $\chi=M/H$ 
at 10~kOe from 2~K to 350~K. 
The susceptibility of \ccro{} has a distinct broad maximum at around 200~K.
In contrast, such a distinct peak is absent in the other samples.
We note that the susceptibility of our samples contain smaller Curie tails at low temperatures 
than those reported by Labeau~\etal\cite{Labeau1980JSolStaChem} and Krimmel~\etal\cite{Krimmel2008PRB}
Among our samples of \ccro{}, there is a systematic tendency of an increase in the susceptibility 
at the broad peak at 200~K as the Curie tail becomes smaller.
In search of possible superconductivity of these metallic compounds, 
we measured the AC susceptibility of all the samples down to 20~mK, 
but found no evidence for any magnetic transitions.


Figure~\ref{fig:f4}(b) shows the specific heat $C_P$. 
There are no indications of magnetic transitions.
The $C_P$-$T$ curves of all the samples are similar 
and $C_P$ seem to approach around 400~J/f.\,u.\,mol\,K at 300~K
(f.\,u.: formula unit). 
These values correspond to 80\% of the value estimated 
from the Dulong-Petit law $C_V=3NR\simeq499$~J/f.\,u.\,mol\,K, 
where $N=20$ is the number of atoms per formula unit, and $R$ is the gas constant.

\begin{figure}[!tb]
\begin{center}
\includegraphics[width=6.50cm,angle=270]{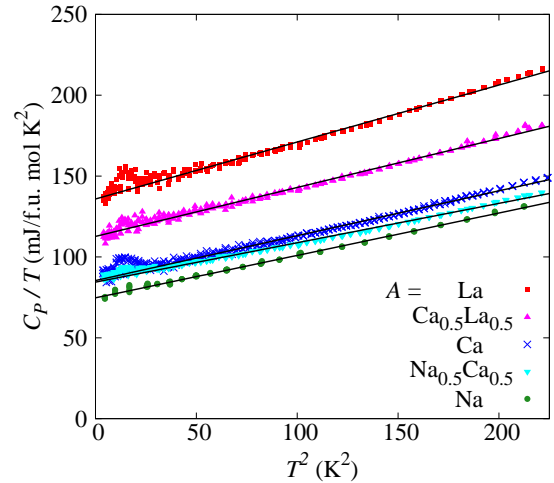}
\caption{(Color online)
Variation of the specific heat of \acro{} divided by temperature, $C_P/T$, 
as a function of $T^2$.
The lines are obtained by fitting the equation 
$C_P/T$=$\gamma{}+\beta{T^2}$ to the data from 5~K to 15~K.
\label{fig:f5}}
\end{center}
\end{figure}

$C_P/T$ as a function of $T^2$ in a low temperature range is shown in Fig.~\ref{fig:f5}.
The solid lines represent the fitting $C_P/T=\gamma{}+\beta{}T^2$ from 5~K to 15~K. 
We note that in this paper the values of $\gamma$ are given per formula unit.
For \ccro{}, for example, $\gamma=85$~mJ/f.\,u.\,mol\,K$^2$ 
corresponds to $\gamma=28$~mJ/Cu\,mol\,K$^2$, 
consistent with the earlier reports\cite{Rami2004SolStaCommun,Koba2004JPSJ,Krimmel2008PRB}.
The values of $\gamma$ of the other samples are also large 
and in particular \lcro{} has the largest $\gamma${}.
In fact, as the \textit{A}-site ions vary from Na\sps{+} to La\sps{3+}, 
$\gamma$ exhibits a systematic increase.
The Debye temperatures $\varTheta\subm{D}$ of all these samples are 
evaluated to be about 500~K using the relation 
$\varTheta\subm{D}={(12\pi^4NR/5\beta)^{1/3}}$.

\begin{figure}[!tb]
\begin{center}
\includegraphics[width=7.00cm,angle=270]{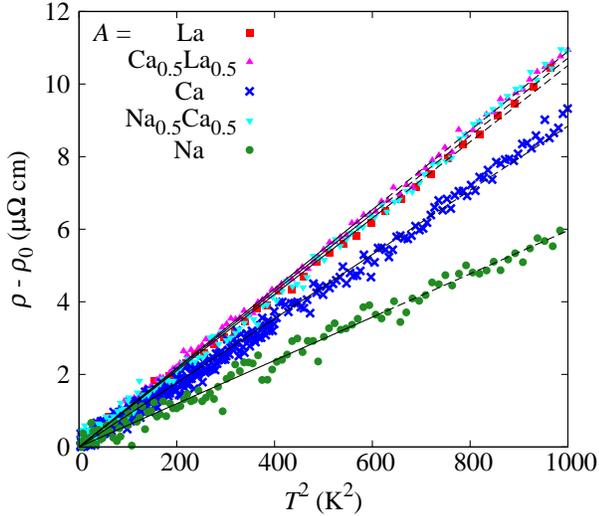}
\caption{(Color online)
Temperature-dependent part of the resistivity $\rho-\rho_0$ plotted against $T^2$ for \acro{}.
\label{fig:f6}}
\end{center}
\end{figure}


Fig.~\ref{fig:f4}(c) represents the resistivity $\rho$ of all the samples.
In order to characterize the temperature dependence at low temperatures, 
we fit the relation $\rho(T)=\rho_0+A'T^{\alpha}$ to the data 
using $\rho_0$, $A'$, and $\alpha$ as the fitting parameters.
The resulting exponents $\alpha$ for all members lies within $2.0\pm0.2$ below 25~K, 
demonstrating the $T^2$ behavior.
This quadratic temperature dependence is also evident in Fig.~\ref{fig:f6}, 
where $\rho-\rho_0$ is plotted against $T^2$.
The coefficients $A$ of the $T^2$ term listed in Table~\ref{table:t1} are evaluated 
by the fitting with $\rho(T)=\rho_0+AT^2$ below 25~K.

\begin{table*}[htb]

\caption{
Parameters characterizing the physical properties of \acro{}.
The $\chi\subm{10\,K}$ values are extracted from the raw data of $M/H$.
$\gamma$, $\beta$, $A$, and $\rho_0$ are obtained by fitting 
$C_P/T=\gamma+\beta{}T^2$ from 5~K to 15~K
and $\rho=\rho_0+AT^2$ up to 25~K. 
Using these parameters, we evaluate: %
Debye temperature $\varTheta\subm{D}$, %
Wilson ratio $R\subm{W}$, %
Kadowaki-Woods ratio $a\subm{KW}\equiv{}A/{\gamma^2}$, 
volumetric electronic specific heat coefficient $\gamma\subm{v}$, 
Hussey's Kadowaki-Woods ratio $a\subm{KWH}\equiv{}A/{\gamma\subm{v}}^2$ 
with $b_0=1$ $\mu${}$\Omega$\,cm\,K$^2$\,cm\sps{6}/mJ$^2$, 
and Jacko's Kadowaki-Woods ratio in three-dimensional systems $a\subm{KWJ}$.
\label{table:t1}}

\begin{tabular}{ccccccc} %
\hline
\textit{A}-site & lattice parameter $a$ & $\chi\subm{10\,K}$ & $\gamma$ & $\beta$ & $A$ & $\rho_0$ \\
ions & (nm) & ($10^{-3}$ emu/f.\,u.\,mol) & (mJ/f.\,u.\,mol\,K$^2$) & 
(mJ/f.u.\,mol\,K$^4$) & (n$\Omega$\,cm/K$^2$) & ($\mu${}$\Omega${}\,cm) \\ %
\hline
La & $0.7477\pm0.0001$ & 6.17 & 136 & 352 & 10.2 & 231 \\
\cl & $0.7460\pm0.0001$ & 4.72 & 113 & 303 & 10.8 & 135 \\
Ca & $0.7421\pm0.0001$ & 4.72 & 85 & 277 & 8.5 & 36 \\
\nc & $0.7407\pm0.0002$ & 3.56 & 84 & 243 & 10.2 & 78 \\
Na & $0.7386\pm0.0001$ & 3.14 & 75 & 262 & 6.1 & 8 \\ %
\hline
\end{tabular}

\begin{tabular}{ccccccc} %
\hline
\textit{A}-site & $\varTheta\subm{D}$ & $R\subm{W}$ & $a\subm{KW}$ & $\gamma\subm{v}$ & $a\subm{KWH}$ & $a\subm{KWJ}$ \\
ions & (K) &  & 
(($\mu${}$\Omega$\,cm/K$^2$)/(mJ/Ru\,mol\,K$^2$)$^2$) & (mJ/cm$^3$\,K$^2$) & 
($\mu${}$\Omega$\,cm\,K$^2$\,cm\sps{6}/mJ$^2$) &  \\ %
\hline
La & 480 & 3.3 & $0.9\times{10^{-5}}$ & 1.08 & 0.009\,$b_0$ & 458 \\
\cl & 505 & 3.1 & $1.3\times{10^{-5}}$ & 0.90 & 0.013\,$b_0$ & 662 \\
Ca & 519 & 4.0 & $1.8\times{10^{-5}}$ & 0.69 & 0.018\,$b_0$ & 845 \\
\nc & 543 & 3.1 & $2.1\times{10^{-5}}$ & 0.69 & 0.021\,$b_0$ & 963 \\
Na & 529 & 3.1 & $1.7\times{10^{-5}}$ & 0.62 & 0.016\,$b_0$ & 684 \\ %
\hline
\end{tabular}
\end{table*}

\section{Discussion}

In this section we examine the origin of the mass enhancement in \acro{}
from the relations among electronic specific heat coefficients $\gamma$, 
the susceptibility values $\chi\subm{10\,K}$, 
and the coefficients $A$ of the $T^2$ term in the resistivity.


We evaluate the Wilson ratio 
$R\subm{W}\equiv{}\pi^2k\subm{B}^2\chi_0/(3\mu\subm{B}^2\gamma$), 
using $\chi_0=\chi\subm{10\,K}$ obtained at 10~K.
Theoretically the value of $R\subm{W}$ equals 1 for a free electron gas, 
whereas it approaches 2 in the strong correlation limit for local Fermi liquids.
It may become greater than 2 if ferromagnetic fluctuations are present.
The size of the deviation from unity is regarded 
as a measure of the strength of many-body electron correlations.
$R\subm{W}$ of the present compounds are all above 3 
and comparable to that of some other heavy-fermion compounds.


The relation between $\gamma$ and $A$ is described 
by the Kadowaki-Woods ratio~\cite{KW} $a\subm{KW}\equiv{}A/{\gamma^2}$, 
which is $1\times10^{-5}$~($\mu$$\Omega$\,cm/K$^2$)/(mJ/mol\,K$^2$)$^2$ 
for some typical heavy-fermion metals.
If $\gamma$ is expressed in terms of mJ/Ru\,mol\,K$^2$,
$a\subm{KW}$ of \acro{} (\textit{A} = Na, Ca, La) are all 
around $1\times10^{-5}$~($\mu$$\Omega$\,cm/K$^2$)/(mJ/Ru\,mol\,K$^2$)$^2$, 
consistent with such universality.
In order to remove the ambiguity in the choice of the unit of $\gamma$, 
Hussey\cite{Hussey} recently proposed the modified definition $a\subm{KWH}\equiv{}A/{\gamma\subm{v}}^2$.
Here a molar quantity $\gamma$ is rescaled to 
the volumetric quantity $\gamma\subm{v}\equiv\gamma{}Z/(N\subm{A}{a^3})$, 
where $Z$ is the number of formula units per unit cell, 
$N\subm{A}$ is Avogadro's number, and $a$ is a lattice parameter. 
The modified ratio for all \acro{} is evaluated as 
$a\subm{KWH}\simeq0.01$--$0.02\,b_0$, 
where $b_0\equiv1$ 
$\mu${}$\Omega$\,cm\,K$^2$\,cm\sps{6}/mJ$^2$. 
We note that the rescaled values for all the present compounds happen to be very similar to 
that of a three-dimensional, 
\textit{d}-electron heavy-fermion-like oxide 
LiV\sub{2}O\sub{4}\cite{Urano2000PRL} ($a\subm{KWH}\simeq0.02\,b_0$).

Following Hussey's work, 
Jacko~\etal\cite{Jacko} very recently proposed 
an extended definition of the Kadowaki-Woods ratio applicable to a wide variety of 
heavy fermions, transition metals, transition-metal oxides, and organic charge transfer salts.
They introduced the dimensionless ratio 
$a\subm{KWJ}\equiv4\pi{}\hbar{}k\subm{B}^2e^2f_{d}(n)(A/{\gamma\subm{v}}^2)$, 
where the function $f_{d}(n)$ is expressed 
in terms of the electron density $n$ and spatial dimensionality $d$ of a system.
They further proposed the universality $a\subm{KWJ}=81$. 
If we assume that \ccro{} contains 16~conduction electrons per formula unit of Ru 4$d^4$, 
the electron density is estimated to be $n=16\,Z/{a^3}$, 
and we obtain $a\subm{KWJ}\simeq880$ 
using $f_3(n)=\sqrt[3]{3n^7/\pi^4\hbar^6}$ for a spherical Fermi surface.
This evaluation is based on a crude single-band picture.
Considering the change in the number of conduction electrons, 
the values of $a\subm{KWJ}$ for the other \acro{} also lies within a factor of 2 (See Table~\ref{table:t1}).
Although these values are an order of magnitude larger than the universal value proposed by Jacko~\etal, 
they are within the range of the values for other highly-correlated electron systems.


As shown in Table~\ref{table:t1}, $\gamma$, 
$\chi\subm{10\,K}$, 
and $A$ exhibit almost systematic increases 
with varying the \textit{A}-site ions from Na\sps{+} to La\sps{3+}.
These parameters are closely related to the density of states $D(\epsilon\subm{F})$: 
$\gamma$ and $\chi\subm{10\,K}$ are expected to be proportional to $D(\epsilon\subm{F})$ 
and $A$ is expected to be proportional to $D^2(\epsilon\subm{F})$ for Fermi liquids. 

Now we attempt to explain these systematic increases first 
from the standpoint of the rigid band model with the total density of states spectra 
obtained by the band calculation of \ccro{} using exchange correlation potential 
(shown in Fig. 3 in Ref. 25).
A rigid band model relies on the approximation 
that the change in the valence of an ion induced by some site substitution results only in 
a shift of $\epsilon\subm{F}$ without changing the band structure.
For \ccro{}, when we substitute Na\sps{+} (La\sps{3+}) for Ca\sps{2+}, 
that is, when the number of the conduction electrons decreases (increases), 
$\epsilon\subm{F}$ would shift to lower (higher) energy in the calculated spectra.
Assuming the rigid band model, $D(\epsilon\subm{F})$ would decrease (increase) 
since the calculated spectra has a positive slope 
around the Fermi energy $\epsilon\subm{F}$; 
$\gamma$, $\chi\subm{10\,K}$, and $A$ would all decrease (increase).
From a crude estimation using the calculated spectra, 
$D(\epsilon\subm{F})$ deviate $\sim\pm25\%$ by substituting Na\sps{+} or La\sps{3+}
(13.4, 17.4, and 22.0 electrons/eV 
for \textit{A} = Na, Ca, and La, respectively). 
These \textit{A}-site substitution results with the rigid band model 
are reasonably consistent with our observation
and support that $\gamma$ of \ccro{} should be mainly enhanced by the band itself.
The present analysis implies that the changes 
in the lattice parameter plays a secondary role in the mass enhancement, 
consistent with the interpretation by Ramirez~\etal\cite{Rami2004SolStaCommun}.


To date it has been proposed 
that the Kondo effect between the localized Cu~3\textit{d} and the itinerant Ru~4\textit{d} electrons 
may be the origin of the heavy effective mass 
since the behavior of the maximum around 200~K of the $\chi$-$T$ curve of \ccro{} is similar to 
that of the \textit{f}-electron heavy fermion compound CeSn\sub{3}~\cite{Misawa1974SolStaCommun}.
We should stress here that indeed we observe 
the peak in $\chi(T)$ ascribable to the Kondo effect only for \ccro{} (See Fig.~4(a)).
However, if the Kondo effect is the dominant origin for the mass enhancement of \acro{}, 
a largest value of $\gamma$ is expected for \ccro{}.
In reality, \ccro{} is not the only one which has a large $\gamma$ value.
The other members, 
in which the formal valence of Cu ions is shifted from (localized) 2+, 
also have large $\gamma$ values. 
It is clear that the Kondo effect cannot be the dominant origin of the mass enhancement of \acro{}.

\section{Conclusion}

We investigated basic physical properties of \acro{ }\Aall{} to low temperatures.
Not only the electronic specific heat coefficients of \ccro{}, 
in which the formal valence of Cu ions is 2+ leading to localized Cu~3\textit{d} electrons, 
but also that of the other members of \acro{} are relatively large.
This fact is inconsistent with the scenario 
that the Kondo effect plays a dominant role for the mass enhancement of \ccro{}.

Nevertheless, since we observed the maximum in $\chi(T)$ only in \ccro{} around 200~K, 
which is similar to the maximum for \textit{f}-electron heavy fermion compounds CeSn\sub{3}, 
the correlation between the localized Cu~3\textit{d} and the itinerant Ru~4\textit{d} electrons (Kondo effect) 
may in part contribute to the mass enhancement of \ccro{} to minor extent.
The Kondo effect derived from the localized Cu\sps{2+} electrons cannot be totally neglected for \ccro{}.

The Wilson ratio and Kadowaki-Woods ratio of the present compounds exhibit 
values similar to those of other heavy-fermion compounds, 
suggesting the mass enhancement of \ccro{} 
originating mainly from the correlation among the Ru~4\textit{d} electrons.
The electronic specific heat coefficient $\gamma$, 
the susceptibility value $\chi\subm{10\,K}$, 
and the $\rho$-$T^2$ coefficient $A$ exhibit almost systematic increases 
with varying the valence of \textit{A}-site ions from 1+ to 3+.
A simple rigid band model can account for our experimental results,
supporting that the heavy mass arising from the Ru bands and the shape of the bands are important. 

More systematic investigation is under progress 
to characterize the precise variations of $\gamma$ for other combinations of the \textit{A}-site.
At the same time it is desirable to determine the valence states of Cu and Ru ions in \acro{} 
to clarify the roles of the localized and itinerant Cu electrons.

\section*{Acknowledgment}
The authors thank Shunichiro Kittaka for experimental support 
and Masaki Azuma, Takaaki Sudayama, and Takashi Mizokawa for valuable discussions.
We also thank Keyence Co. for the optical microscope 
used to take images of crystals of \ncro{}. 
This work has been supported by the Grant-in-Aid 
for the Global COE Program "The Next Generation of Physics, Spun from Universality and Emergence" 
from the Ministry of Education, Culture, Sports, Science and Technology (MEXT) of Japan.
It has also been supported by Grants-in-Aid for Scientific Research from MEXT 
and from the Japan Society for the Promotion of Science.


\end{document}